# Artificial Intelligence to Enhance Mission Science Output for In-situ Observations: Dealing with the Sparse Data Challenge

M. I. Sitnov, G. K. Stephens, V. G. Merkin, C.-P. Wang, D. Turner, K. Genestreti, M. Argall, T. Y. Chen, A. Y. Ukhorskiy, S. Wing, Y.-H. Liu

The goal of space exploration is to better understand the universe encompassing our small planet and its life. Space research includes remote (similar to observational astronomy) and in-situ observations made at the immediate location of the spacecraft. Observations of the first type usually deal with high-resolution and high cadence data (e.g., of the Sun's disc images), often referred to as the so-called **Big Data** (e.g., Armstrong and Fletcher, 2019). In the last decade there has been great progress with processing this data due to a revolution in machine learning provided by a new generation of multi-layer artificial neural networks (e.g., LeCun et al., 2015). In this WP we would like to draw the community's attention to another challenge of **Sparse Data** sampling in case of in-situ observations (e.g., Sitnov et al., 2020 and refs. therein). For instance, in the case of the Earth's magnetosphere, there are fewer than a dozen dedicated probes beyond the low-Earth orbit available for magnetospheric observations at any given time. As a result, we still poorly understand the global structure and evolution of the magnetosphere, its magnetic field, electric currents, plasma pressure and high-energy plasma populations. We still don't understand the mechanisms of the main activity processes, magnetic storms and substorms (McPherron, 2016; Wolf et al., 2017). The 2013 Heliophysics Decadal Survey called for a future Magnetospheric Constellation (MagCon) mission consisting of dozens of spacecraft to address such data sparsity challenges in such an enormous volume of space. Considering such future missions and the inescapable data sparsity challenge, along with the advance of instruments, orbits and probes, it is also critical to advance the application of **Artificial Intelligence (AI)** technologies to the analysis of space physics datasets, such as machine learning (ML), data mining (DM) and data assimilation (DA) (e.g., LeCun et al., 2015; Kubat, 2015; Kalnay, 2006).

**Some examples of what has been achieved so far**

K-nearest neighbors data mining (KNN DM) combined with flexible and extensible magnetic field architectures helped organize storm (Tsyganenko & Sitnov, 2007; Sitnov et al., 2008) and substorm (Stephens et al. 2019; Sitnov et al., 2019) magnetometer measurements to reconstruct the structure and morphology of the ring current and substorm-time thinning of the current sheet and dipolarization processes in the magnetotail. The concept of the DM approach is illustrated in Fig. 1: A large historical database of magnetometer measurements is mined to select a small subset of data (Fig. 1b) acquired at times similar to the event of interest in terms of global activity parameters, such as the *Sym-H* index and its time derivative (Fig. 1a). The subset, which is much larger than the few actual in-situ satellites for any given event, can be used to fit a very flexible magnetic field architecture and reveal details of the magnetosphere such as the formation of new X-lines (Fig. 1c).

Similar reconstructions of the substorm growth-phase distributions of the plasma sheet pressure have been obtained using another ML tool, Support Vector Regression Machine (SVRM) (Yue et al., 2015). Artificial neural networks (ANN) have been used to reconstruct the plasmaspheric density and distributions of waves that control particle acceleration and loss in the inner magnetosphere (Bortnik et al., 2018). Shprits et al. (2013) proposed to use DA with Kalman filters to adjust the equations of the particle diffusion in radiation belts using CRRES observations. Merkin et al. (2019) used the empirical pressure distributions derived from the DM magnetic field reconstructions to adjust the MHD equation of state in global modeling of magnetic storms. Wing et al. (2016) used information theory to untangle the solar wind drivers of the radiation belt electrons. Argall et al. (2020) used ANNs to optimize burst-mode data selection in MMS observations, and in particular, to recognize potential reconnection regions at the magnetopause.

**What is desired/proposed**

**Advancing datasets:** Extending the regions and data types involved in the AI. This includes:

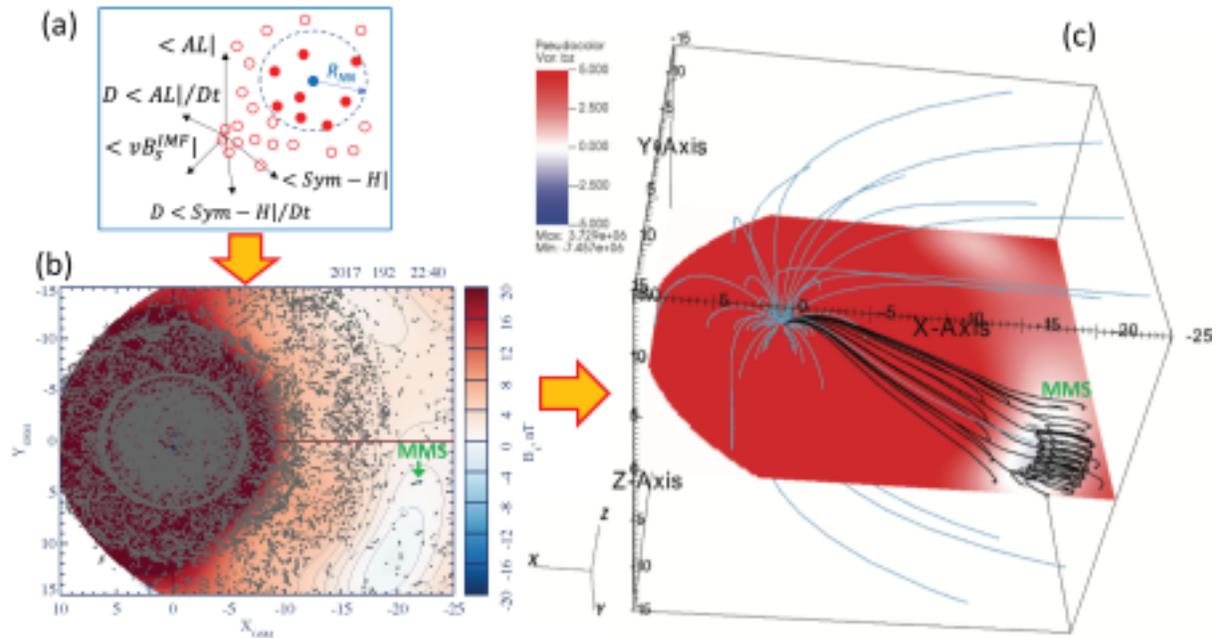

**Fig. 1**. The scheme showing reconstruction of the global shape of the substorm magnetosphere in terms of the geomagnetic field using KNN DM method (Stephens et al., 2019; Sitnov et al., 2019): (a) selecting nearest neighbors for the event of interest in the global parameter space; (b) finding the corresponding subset in the magnetic field database; (c) the resulting magnetic field reconstruction (here for the 11 July 2017 MMS EDR event (Torbert et al., 2017)).

1. Filling spatial gaps in the Earth's magnetosphere, and in particular, the nightside magnetosphere in the region $-55R_E < x < -31R_E$,
2. Processing magnetometer data for other planets (e.g., Saturn and Mercury).
3. Using data from existing ground-based and LEO facilities, such as SuperMag and Iridium/AMPERE. 4. Using heterogeneous data sets that combine different parameters, such as the magnetic field, plasma density, temperature, bulk flow velocity, high-energy plasma intensity etc.

**Advancing AI algorithms:** The Heliophysics community should elaborate or implement new AI algorithms, such as the Long Short-Term Memory ANNs and distance-weighted KNN DM. Elaborate new data ingestion and assimilation techniques. Information Theory can be used to optimize DM, ML and DA methods, and in particular, optimize global parameters of the solar wind and magnetospheric activity to be used as inputs for AI models.

**Empirical modeling of Geospace and beyond to understand the underlying activity mechanisms:** Use AI models to attack the most compelling science questions, such as the mechanisms of magnetic storms, substorms, and the complexity of global magnetospheric convection and energy transport, steady and unsteady collisionless reconnection, coronal heating and solar wind turbulence.

**Predicting Geospace weather using AI:** The Heliophysics community should use ML models to forecast solar activity, the resulting solar wind and IMF perturbations, as well as the distributions of the key parameters in the Geospace environment, such as the geomagnetic field, plasma moments, high-energy particle intensities (including radiation belts), geomagnetically induced currents (GICs) and spacecraft charging.

**Combining AI and first-principle models:** The Heliophysics community should elaborate on the ingestion of the empirical plasma pressure to global MHD models for magnetic storms and substorms. Some related goals include: to combine global reconstructions of storms and substorms with regional kinetic (particle-in-cell or hybrid) simulations; to merge ionosphere/thermosphere models with the corresponding empirical data sets and ML/DM models; to advance AI-enabled GIC models combining global MHD simulations and observations.

**AI-enabled missions:** The Heliophysics community should investigate the use of AI in selecting data from Constellation-class missions. Design special AI-enabled missions, where, for example, a few probes fill a substantial region in space and then used as a proxy of the >1000-probe-equivalent mission empowered by KNN DM algorithms. In particular, our estimates show that the use of 3 probes for >1 year in the tail gap region -$55R_E<x<-31R_E$ enables a cloud of point measurements, which can be sorted using KNN DM to create an empirical picture of that region with the spatial resolution ~1-2RE, temporal resolution ~15 min and variability capturing substorm reconfigurations of the magnetotail. These data can be used to describe the key processes in this heavily under-sampled region of Earth's magnetosphere, which acts like a magnetospheric watershed where the solar wind plasma enters the magnetotail and stored energy is diverted either toward the inner magnetosphere, where it contributes to global magnetospheric convection or tailward, which ultimately results in energy and plasma escaping the magnetosphere.